\RequirePackage{lineno}
\RequirePackage{graphicx,amssymb,array,tabularx,url,xspace}

\documentclass[a4paper]{jpconf}
\usepackage{graphicx}

\usepackage{cite}

\usepackage{amsmath}
\usepackage{rotating}
\usepackage{subfigure}
\usepackage{multirow}

\usepackage{color}\usepackage{soul}

\usepackage{grffile}

\makeatletter{}
\newcommand{\degree}{^{\rm o}}

\newcommand{\PbPb}{Pb--Pb\xspace}

\newcommand{\GeV}{GeV/$c$\xspace}

\newcommand{\sNN}{$\sqrt{s_{\mathrm{NN}}}$}

\begin{document}
\title{Jet-hadron correlations relative to the event plane at the LHC with ALICE}

\author{Joel Mazer}

\address{Department of Physics, University of Tennessee, Knoxville, Tennessee 37996-1200, USA}

\ead{joel.mazer@cern.ch}

\begin{abstract}
Jet quenching is observed at both RHIC and LHC energies.  This suggests  that 
partons lose energy as they traverse the medium.  When the trigger jet is 
studied relative to the event plane, the path length dependence of medium 
modifications can be studied.  We present measurements of the angular correlations relative to the 
event plane between reconstructed jets and charged hadrons in Pb--Pb collisions at 
$\sqrt{s_{\mathrm{NN}}}=$2.76 TeV in ALICE.  A newly implemented, robust background subtraction method 
to remove the complex, flow dominated, combinatorial background is used in this analysis. 
\end{abstract}

\section{Introduction}\label{Sec:Intro}
Jets are ideal probes of the Quark Gluon Plasma (QGP) because 
they originate from hard-scattered partons created early in the collision, prior to the formation 
of the medium.  These partons are modified in the presence of a medium through collisional energy 
loss and induced gluon radiation.  This modification is observed at both LHC and RHIC energies via 
the suppression of high-momentum particles \cite{Adare:2007vu,Adams:2003kv,Adare:2010mq,Adam:2015ewa}.  
A suppression is also seen for high $\it{p}_{T}$ di-hadron 
correlations \cite{PhysRevLett.95.152301,PhysRevC.80.064912,PhysRevC.85.014903,PhysRevC.82.024912}.  

By reconstructing jets, ideally, a more complete picture of how lost energy is redistributed and how 
the effects of jet quenching emerge can be obtained.  This article will discuss the 
status of azimuthal correlations of reconstructed jets binned relative to the event plane 
with charged hadrons in ALICE.

\section{Experiment}\label{Sec:ExpSetup}
Clusters measured in the Electromagnetic Calorimeter (EMCal) and charged hadrons measured with the 
central tracking system allow ALICE to study fully reconstructed jets \cite{Adam:2015ewa}.  
Charged hadrons are reconstructed from their tracks using information from the Time 
Projection Chamber (TPC) and the Inner Tracking System (ITS).

Tracks are reconstructed over the full azimuthal range and come from mid-rapidity in a range of 
$|\eta_{lab}|<$0.7 when reconstructed in a jet, and $|\eta_{lab}|<$0.9 for the associated charged hadrons.  
The EMCal has an acceptance window of $|\eta_{lab}|<$0.7 
and $|\Delta \phi|$=107$\degree$.  For a complete description of the ALICE detector, see 
\cite{Abelev:2014ffa,Abeysekara:2010ze,Abelevetal:2014dna,Alme2010316}.

\section{Jet-hadron correlations}\label{Sect:jethadcorr}

\subsection{Data sample}\label{subsect:Datasample}
The data used for the correlation analysis was collected by the ALICE Experiment in 2011 during 
the 2.76 TeV Pb--Pb collision data taking.  This work uses events which fired the gamma trigger used 
by the ALICE EMCal \cite{Abeysekara:2010ze}.  In addition, a cluster constituent of the trigger jet was required 
to be matched to the fired trigger patch of the event.

\subsection{Jet Reconstruction}\label{subsect:JetReco} 
Jets are reconstructed with a resolution parameter of R=0.2 using the anti-$k_{T}$ jet-finding 
algorithm from the FastJet package \cite{Cacciari:2011ma}.  Jets reconstructed in this analysis 
require a leading cluster to have $E_{T}>$6.0 GeV, which exceeds the trigger threshold.  
The jets are reconstructed with constituent tracks of 
$\it{p}_{T}>$3.0 \GeV and clusters of $E_{T}>$3.0 GeV.  Since the primary goal of jet finding 
is to reconstruct the jet axis accurately, this high constituent cut limits the influence of 
background on jet finding.

\subsection{Measurement}\label{subsect:Measurement}
We define the jet-hadron correlation function in heavy-ion collisions by Eq.~\ref{correlation_eqn_PbPb}.

\setlength{\abovedisplayskip}{-5pt}
\setlength{\belowdisplayskip}{5pt}

\begin{equation}
 \frac{1}{N_{trig}} \frac{d^2N_{assoc,jet}}{d\Delta\phi d\Delta\eta} = \frac{1}{\epsilon a N_{trig}} 
 \frac{d^2 N^{same}_{pairs} }{d\Delta\phi d\Delta\eta}
 -b_0(1+ \sum v^{trig}_n v^{assoc}_n \cos(n\Delta \phi))
 \label{correlation_eqn_PbPb}
 \end{equation}

\noindent Here the first term represents the same event pairs which are divided by an acceptance 
correction, a, provided by mixed events.  Mixed events are used 
to correct for the limited detector efficiency and acceptance for track pairs.  
The single track reconstruction efficiency of associated particles is denoted by $\epsilon$.  
The second term of Eq.~\ref{correlation_eqn_PbPb} is the combinatorial heavy-ion 
background where $b_{0}$ is the background level and the $v_n$ terms are the Fourier 
coefficients of the trigger jet and associated particles.

The trigger jets in this analysis are binned in angle relative to the event plane to 
explore the path length dependence of medium modifications.  The orientations are defined 
such that in-plane is $0 < | \Delta \varphi | < \frac{\pi}{6}$, mid-plane 
is $\frac{\pi}{6} < | \Delta \varphi | < \frac{\pi}{3}$, and out-of-plane 
is $\frac{\pi}{3} < | \Delta \varphi | < \frac{\pi}{2}$, where $\Delta\varphi$ 
denotes the angular difference between the trigger jet and the reconstructed event plane.

\subsection{Background subtraction}\label{subsect:backgroundsub}
When the trigger jet is restricted relative to the event plane, both the background 
level and $v_{n}$ are modified and will contain a dependence on 
the reaction plane resolution $\Re$.  The derivation of the event plane dependent background 
equations are given in \cite{Bielcikova:2003ku}.  The reaction plane 
resolution corrects for the difference 
between the reconstructed event plane and the underlying symmetry plane, $\Psi_{n}$.

Various background subtraction methods were developed in 
\cite{Sharma:2015qra} and applied to the azimuthal correlation functions.  The primary 
method is the reaction plane fit (RPF).  The 
RPF method works under the assumption that the signal is negligible 
in the large $\Delta \eta$ and small $\Delta\phi$ region.  The 2D correlations are projected over 0.8$<|\Delta\eta|<$1.2 
to define the background dominated region, while the signal+background region is defined to be $|\Delta\eta|<$0.6.  
To ensure as much information as possible is going into the fit by constraining the shape and level of the background, 
the in-plane, mid-plane, and out-of-plane orientations are simultaneously fit up to fourth order in $v_{n}$ 
and required to have the same fit parameters.

\begin{figure}[!htbp]
\centering 
\begin{minipage}[c]{37pc}
\centering \includegraphics[width=37pc]{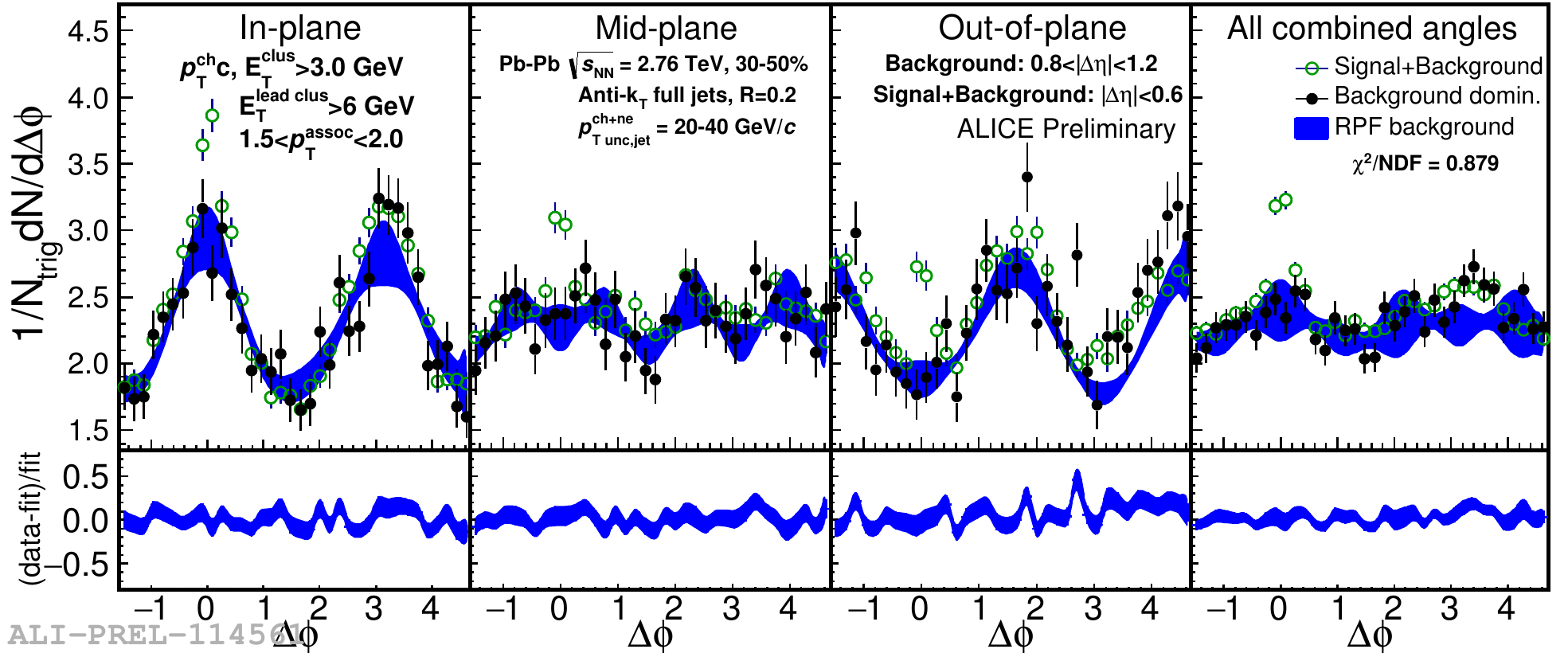}\hspace{1pc}\end{minipage}
\caption{\label{fig:RPFfitData1.5-2.0} Signal+background region, background dominated region and RPF fit to background 
are shown for the 1.5-2.0 \GeV associated $\it{p}_{T}$ bin for full jets in \PbPb collisions 
at \sNN{} = 2.76 TeV for the 30-50\% most central events.  The blue band shows the uncertainty of 
the background fit, which is non-trivially correlated point-to-point \cite{Sharma:2015qra,Nattrass:2016cln}.}
\end{figure}

The background dominated region is fit over the region $|\Delta\phi|<\pi/2$, 
shown in Fig.~\ref{fig:RPFfitData1.5-2.0}.  We see from the blue 
band that the RPF fit models the data quite well, even at low $\it{p}_{T}$ where the background is large.  
The higher order Fourier coefficents clearly matter, as 
the background for trigger jets mid-plane has four peaks, consistent with a $v_4$ dependence.  
This method does not require independent measurements of $v_{n}$ and is able to extract the signal with smaller 
errors while requiring fewer assumptions and less bias than prior subtraction methods.  This is especially 
useful since there are currently no $v^{jet}_3$ or $v^{jet}_4$ measurements.

\setlength{\intextsep}{2pt}
\setlength{\belowcaptionskip}{-2.5pt}

\subsection{Results}\label{subsect:Results}
\begin{figure}[!htbp]
\centering 
\begin{minipage}[c]{37pc}
\centering \includegraphics[width=37pc]{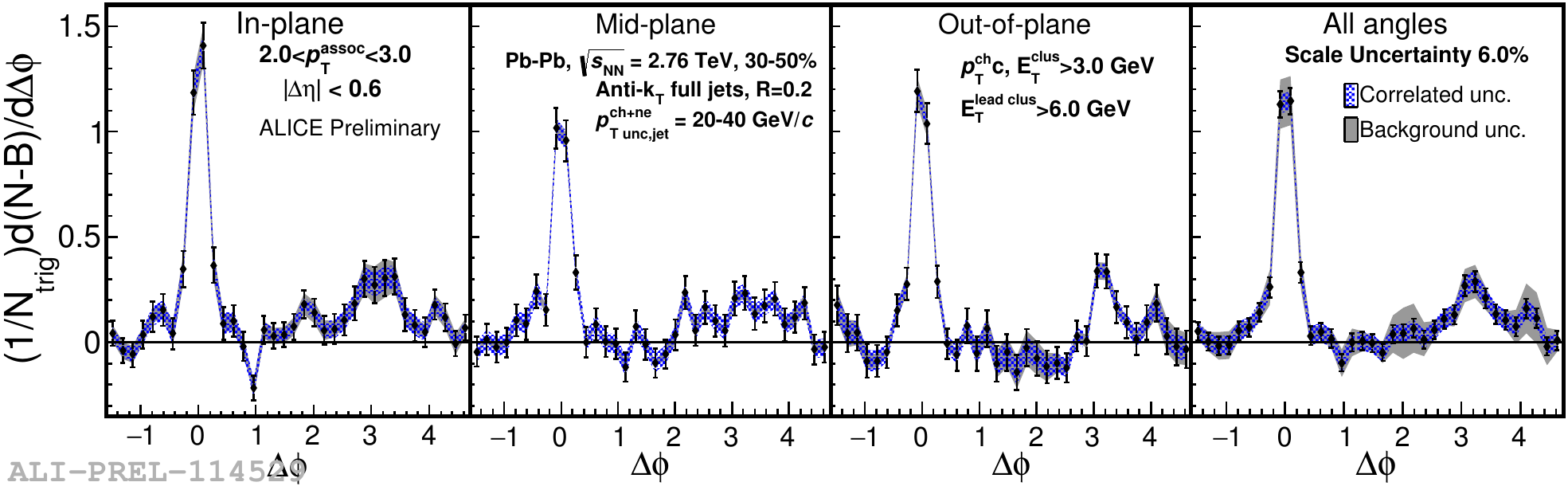}\hspace{1pc}\end{minipage}
\caption{\label{fig:CorrelationPlot2.0-3.0} Corrected $\Delta \phi$ correlation function for 2.0$<\it{p}_{T}<$3.0 \GeV 
associated hadrons for full jets in \PbPb collisions at \sNN{} = 2.76 TeV for the 30-50\% most central events. 
The blue band corresponds to the correlated uncertainty and the grey band corresponds to the 
background uncertainty which is non-trivially correlated point-to-point \cite{Sharma:2015qra,Nattrass:2016cln}.  
There is an additional 6\% global scale uncertainty.}
\end{figure}

We extract the signal by subtracting the large correlated background from the correlation function.  
Figure~\ref{fig:CorrelationPlot2.0-3.0} shows the signal for associated particles of 2.0-3.0 \GeV.  
The uncertainties are dominated by statistics.  With higher statistics, the uncertainties 
could be vastly reduced to allow for a more precise measurement.

We define the yields by Eq.~\ref{eqn:Yield}. 

\begin{equation}
 Y = \frac{1}{N_{trig}} \int_{c}^{d} \int_{a}^{b} \frac{d(N_{meas} - N_{bkgd})}{d\Delta\phi}  d\Delta\phi d\Delta\eta 
 \label{eqn:Yield}
\end{equation}

\begin{figure}[!htbp]
\centering 
\begin{minipage}[c]{18.5pc}
\centering \includegraphics[width=18.5pc]{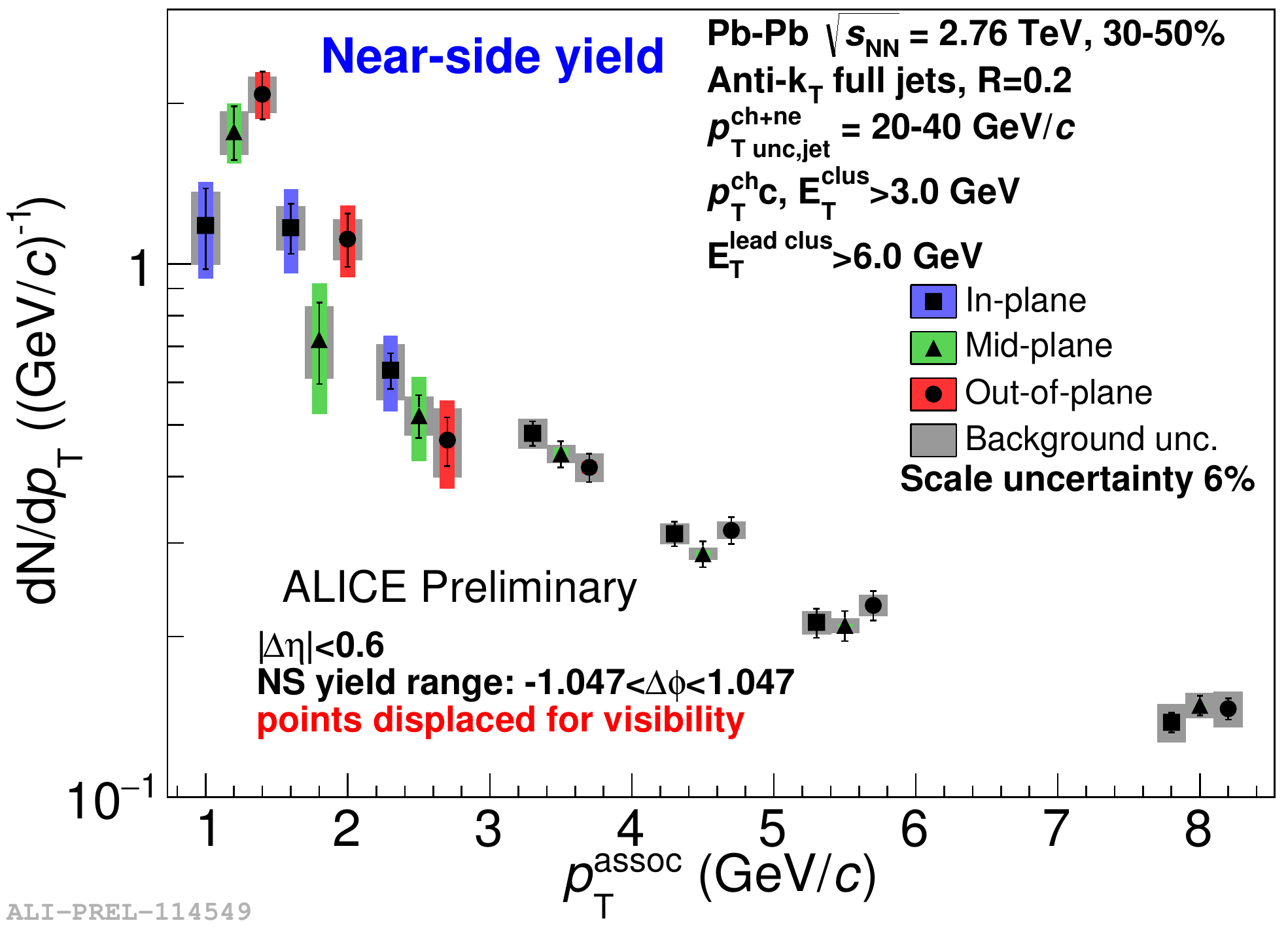}\hspace{1pc}\end{minipage}
\begin{minipage}[c]{18.5pc}
\centering \includegraphics[width=18.5pc]{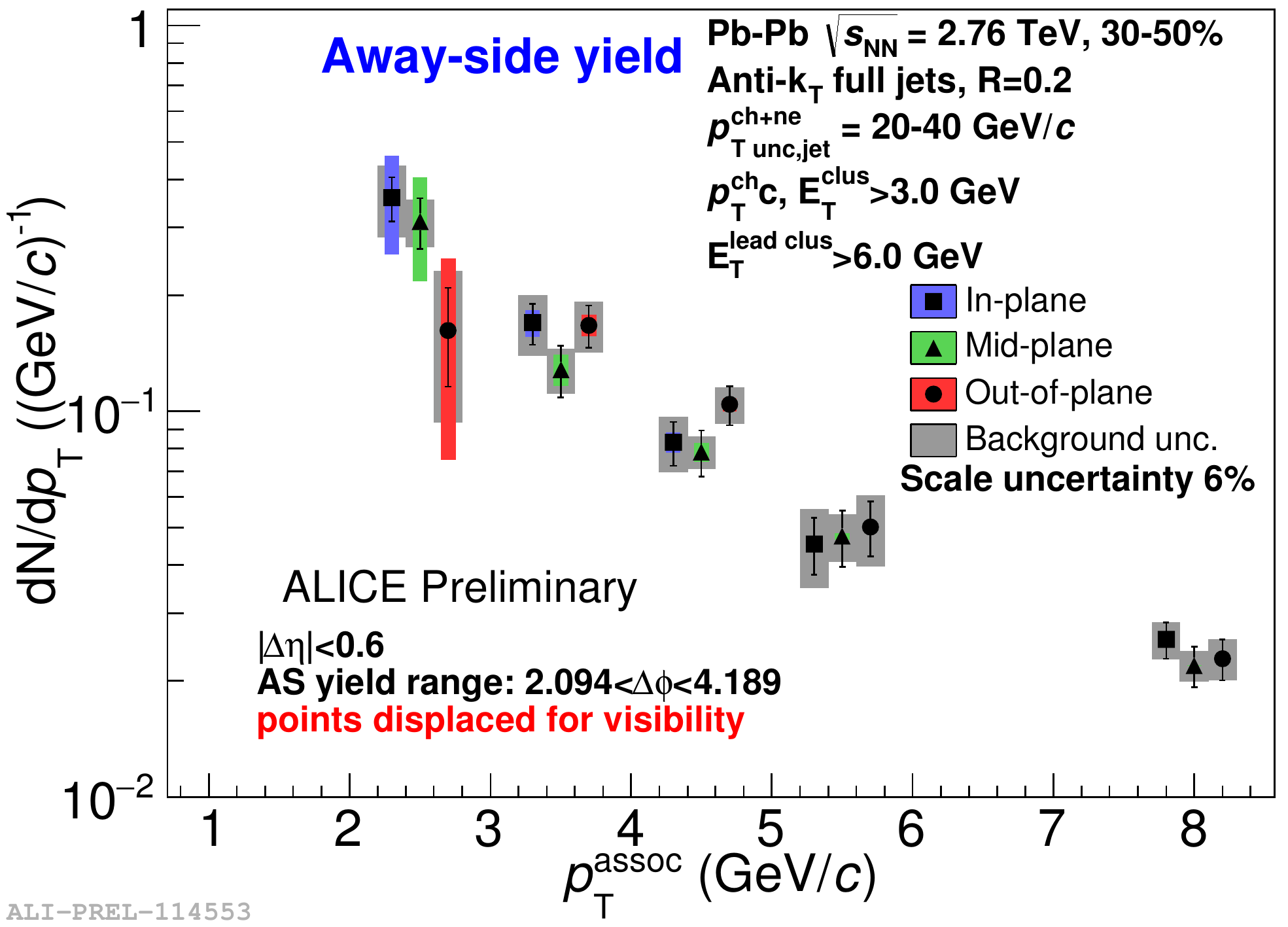}\hspace{1pc}\end{minipage}
\caption{\label{fig:Yield} The near-side yield (left) and away-side yield (right) for full jets 
in \PbPb collisions at \sNN{} = 2.76 TeV for the 30-50\% most central events. The colored bands correspond  
to the correlated uncertainties and the grey band corresponds to the background uncertainty which is 
non-trivially correlated point-to-point \cite{Sharma:2015qra,Nattrass:2016cln}.  There is an 
additional 6\% global scale uncertainty. }
\end{figure}

\noindent Where the integration limits a and b correspond to $\Delta \phi$ values of -1.047 and +1.047 on the 
near-side and +2.094 and +4.189 on the away-side respectively.  In addition, the integration limits c and d correspond 
to $\Delta \eta$ values of -0.6 and 0.6 for both the near-side and away-side.  
The near-side (left) and away-side (right) jet yields for 1.0$<\it{p}^{assoc}_{T}<$10.0 \GeV are 
shown in Fig.~\ref{fig:Yield}.  The near-side yield is consistent with little or no modification.  
There is no clear dependence of the away-side peaks on orientation relative to the event plane.  
There are competing effects across different $\it{p}_{T}$ ranges.  Jet quenching could cause a decrease in yield 
going from in-plane to out-of-plane, while gluon radiation could cause an increase.  

\section{Summary and Outlook}\label{Sec:Summary}
The jet-hadron correlation results were seen to have their uncertainties dominated by statistics.  
No significant event plane dependence was seen to within the current uncertainties on the extracted 
jet yield.  The RPF background subtraction method applied to data was seen to have various advantages in that it does 
not require independent $v_n$ measurements, has less bias and fewer assumptions than prior background 
subtraction methods, and can extract the signal with great precision.  
The current results are consistent with the re-analysis of STAR data seen in \cite{Nattrass:2016cln}.

\section*{References}
\makeatletter{}\providecommand{\newblock}{}

\end{document}